\documentclass[a4paper,12pt]{article}
\usepackage{amsmath}
\usepackage{amssymb}
\usepackage{amsthm}
\usepackage{latexsym}
\usepackage{mathrsfs}
\usepackage[dvips]{graphicx}
\usepackage{pstricks}
\usepackage{color}
\def\f(#1){{\mathop{f}^{(#1)}}}
\def\m(#1){{\mathop{m}^{(#1)}}}
\def\C(#1){{\mathop{C}^{(#1)}}}
\def\p(#1){{\mathop{p}^{(#1)}}}

\def\ben{\begin{equation}}
\def\een{\end{equation}}
\def\bena{\begin{eqnarray}}
\def\eena{\end{eqnarray}}
\def\non{\nonumber}

\def\d{{\rm d}}

\def\M{{\mathbb M}}
\def\C{{\cal C}}

\def\D{{\rm d}}

\def\mr{{\mathbb R}}

\def\B{{\mathscr B}}

\newcommand{\half}{\frac{1}{2}}

\newcommand{\mz}{{\mathbb Z}}

\renewcommand{\H}{{\mathscr H}}

\renewcommand{\M}{{\mathscr M}}

\newcommand{\e}{{\rm e}}

\theoremstyle{theorem}

\newtheorem{thm}{Theorem}

\newtheorem{lemma}{Lemma}

\begin{document}

\title{Horizon area-angular momentum inequality in higher dimensional spacetimes}

\author{
Stefan Hollands$^{1}$\thanks{\tt HollandsS@Cardiff.ac.uk}\:
\\ \\
{\it ${}^{1}$School of Mathematics, Cardiff University} \\
{\it Cardiff, United Kingdom}
}

\date{26 October 2011}
\maketitle

\begin{abstract}
We consider $n$-dimensional spacetimes which are axisymmetric--but not necessarily
stationary (!)--in the sense of having isometry group $U(1)^{n-3}$, and which satisfy
the Einstein equations with a non-negative cosmological constant. We show that any black hole horizon
must have area $A \ge 8\pi |J_+ J_-|^\half$, where $J_\pm$ are
distinguished components of the angular momentum corresponding to linear combinations
of the rotational Killing fields that vanish somewhere on the horizon. In the case of $n=4$, where there is only one
angular momentum component $J_+=J_-$, we recover
an inequality of 1012.2413 [gr-qc]. Our work can hence be viewed as a
generalization of this result to higher dimensions. In the case of $n=5$ with horizon
of topology $S^1 \times S^2$, the quantities $J_+=J_-$ are the same angular momentum component (in the $S^2$ direction).
In the case of $n=5$ with horizon topology $S^3$, the quantities $J_+, J_-$ are the distinct components of
the angular momentum. We also show that,
in all dimensions, the inequality is saturated if the metric is  a
so-called ``near horizon geometry''. Our argument is entirely quasi-local, and hence also applies e.g.
to any stably outer marginally trapped surface.
\end{abstract}


\sloppy

\section{Statement of the result and basic definitions}

In \cite{dain,dain1,dain2}, a remarkable inequality was shown to hold between the area $A$ of a black hole horizon
(or more generally, stably outer marginally trapped horizon), and the angular momentum $J$. The inequality holds in any axisymmetric--but not necessarily stationary (!)--spacetime
of dimension $n=4$ satisfying the vacuum Einstein equations, possibly including a non-negative cosmological constant,
and it states that
\ben
A \ge 8\pi |J| \ .
\een
Although it is not clear whether such an inequality will hold in general, non-axisymmetric spactimes, it is still of considerable interest,
because it is a universal bound for a very wide class of--possibly highly dynamical (!)--spacetimes, which are
very difficult to analyze within any existing scheme. Furthermore, it gives a new perspective upon the Kerr-solution, which is in this class and which is
in addition stationary. In fact, it was shown in~\cite{dain2} that the unique vacuum spacetime saturating the bound
is the so-called ``extremal Kerr near horizon geometry'', which can be viewed as the close-up view of the
geometry near the horizon of an extremal Kerr black hole.

The purpose of the present note is to provide an analogue of this statement in higher dimensions. For this,
we will consider a subclass of $n$-dimensional spacetimes with a ``comparable amount of symmetry'' as axisymmetric
spacetimes in $n=4$ spacetime dimensions. The class we consider consists of those spacetimes $(\M,g)$ which satisfy
the vacuum Einstein equation
\ben
R_{\mu\nu}-\half g_{\mu\nu} R = -\Lambda g_{\mu\nu} \ , \qquad \Lambda \ge 0
\een
with a non-negative cosmological constant,
and which additionally admit an action of $U(1)^{n-3}$ by isometries. We refer to these spacetines again as
axisymmetric. Thus, loosely speaking,
the metric $g$ can depend non-trivially on three coordinates, one of which is time.
The commuting Killing vector fields generating
the isometry group are denoted by $\xi_i, i=1,...,n-3$. One then defines
\ben\label{komar}
J_i = \frac{1}{8\pi} \int \star \D \xi_i \ ,
\een
where the integral is over any closed $(n-2)$-dimensional surface co-bordant to $\B$.
If the spacetime is asymptotically Kaluza-Klein (or asymptotically flat in $n=4,5$),
then this surface can be taken to be at infinity, and the $J_i$ then coincide with the usual
ADM-type conserved quantities with the physical interpretation of angular momenta. Although our analysis is entirely ``quasi-local'', i.e.
involves only the geometry in a vicinity of $\B$, not the asymptotics of the spacetime,
one of course has this case in mind in physical situations.

We next explain what we mean by a {\em stably outer (marginally) trapped surface} (``horizon''). Consider an $(n-1)$-dimensional embedded
null hypersurface $\H$ ruled
by affinely parameterized null geodesics with future directed tangent $n$, and expansion
$\theta_n$ on $\H$. If we have a cross section $\B$ of $\H$ transverse to these
geodesics, we can define a second future directed null vector $l$, perpendicular to $\B$, having the property that $g(n,l)=-1$ on $\B$.
For $\B$ to be a stably outer (marginally) trapped surface it is  required that $\theta_n = 0$, and that ${\mathscr L}_l \theta_n  \le 0$,
on $\B$, and it is required that $\B$ is a compact closed submanifold of dimension $n-2$, so that $-l$ 
is pointing outward. These conditions express that, on $\B$, lightrays tangent to $\H$ are non-focussing,
and become expanding ``slightly outside of $\H$'', and contracting ``slightly inside of $\H$''.
A special case of this setup is furnished by the horizon of a stationary black hole, but
we emphasize that our spacetimes are {\em not} assumed to be stationary in general. Our proof still works if
the first condition is relaxed to $\theta_n \ge 0$ provided we additionally have
$\theta_l \le 0$, but it would not work for a negative cosmological constant $\Lambda<0$.

The cross section $\B$ can be chosen so that it is invariant under the action of $G=U(1)^{n-3}=T^{n-3}$, i.e.
that the mutually commuting Killing vectors $\xi_i$ are all tangent to $\B$. Thus, $\B$ is a compact
$(n-2)$ dimensional manifold with an action of an $(n-3)$ dimensional torus. It is not too difficult to
classify such actions and $\B$'s topologically~\cite{hy}; the only possibilities are:
\bena
\B \cong \begin{cases}
S^3 \times T^{n-5} \, ,\\
S^2 \times T^{n-4} \, ,\\
L(p,q) \times T^{n-5} \, ,\\
T^{n-2} \, .
\end{cases}
\eena
The symbol $L(p,q)$, with $p,q$ mutually prime integers, denotes
a lens space. Furthermore, the orbit space $\hat \B := \B/G$, consists of
a closed interval $\hat \B \cong [-1,+1]$ in the first three cases, whereas
it is $\hat \B \cong T^1 = S^1$ in the last case. The last topology
type is actually inconsistent with the the vacuum field equations and
our geometric conditions on $\B$. In fact, as shown in~\cite{racz,galloway},
these conditions imply that $\B$ can carry of metric of positive scalar curvature,
which $T^{n-2}$ cannot. Thus, we only need to focus on the first three topology types,
in which case we denote by $x \in [-1,+1]$ a coordinate parameterizing the orbit space. One then
shows~\cite{hy,hy1} that there
are integer vectors $a_\pm \in \mz^{n-3}$ such that
\ben
a_\pm^i \xi_i \to 0 \ , \quad \text{at $x=\pm 1$.}
\een
Stated differently, the Gram matrix
\ben\label{gram}
f_{ij} = g(\xi_i, \xi_j)
\een
is non-singular (positive definite) in the interior of the interval and it has a one-dimensional null-space
at each of the two end points, spanned by, respectively, $a_\pm$:
\ben\label{bndycond}
f_{ij}(x) a_\pm^i \to 0 \quad \text{at $x=\pm 1$.}
\een
The integers $a^i_\pm$, which may without loss of generality be assumed to
satisfy ${\rm g.c.d.}(a^i_+) = 1 = {\rm g.c.d.}(a^i_-)$, determine the topology of $\B$ (i.e. which of the first three cases
we are in). Up to a globally defined redefinition\footnote{This
corresponds to an automorphism of $G=T^{n-3}$ and is hence merely
a redefinition of the group action on $\M$.} of the Killing fields of the form
\ben\label{redef}
\xi_j \mapsto \sum A^i_j \xi_i \,\, , \quad A \in SL(\mz, n-3) \, ,
\een
we have
\ben\label{apm}
a_+ = (1,0,0,\dots,0) \, , \quad a_- = (q, p, 0,\dots,0) \, , \quad p,q \in \mz \, , \quad {\rm g.c.d.}(p,q) = 1\, .
\een
It is then simple to see that the topology of $\B$ is, respectively~\cite{hy,hy1}
\ben
\B \cong
\begin{cases}
S^3 \times T^{n-5} & \text{if $p=\pm 1,q=0$,} \\
S^2 \times T^{n-4} & \text{if $p=0, q= \pm 1$,}\\
L(p,q) \times T^{n-5} & \text{otherwise.}
\end{cases}
\een
We are now in a position to state our inequality.
\begin{thm}\label{thm1}
Let $(\M,g)$ be a spacetime satisfying the vacuum Einstein equations with a cosmological constant
$\Lambda \ge 0$ having isometry group at least $T^{n-3}$. Define
\ben
J_\pm := J_i a_\pm^i \ .
\een
Then:
\begin{enumerate}
\item The area of any stably outer marginally trapped surface (e.g. event horizon
cross section of a black hole) satisfies
\ben
A \ge 8\pi |J_+ J_-|^\half \ .
\een
\item
Furthermore, if $\Lambda=0$, and if $(\M, g)$ is a ``near horizon geometry'', then
the inequality is saturated. Conversely, if the inequality is saturated, then
the tensor fields $\alpha, \beta,\gamma$ determining the induced geometry on $\H$
[see eq.~\eqref{gnc}] are equal to those of a near horizon geometry.
\end{enumerate}
\end{thm}
The near horizon metrics referred to in 2) have isometry group $O(2,1) \times G$, and were given
explicitly in~\cite{hi}. Their definition and properties are recalled in the appendix for convenience.

The proof of the theorem is given in section~\ref{proof}. It consists of the following steps: First,
the stably outer marginally trapped condition on $\B$ is used in combination with Einstein's equations to
derive an inequality between various geometric quantities on $\B$, which are integrated over $\B$
against a testfunction, see lemma~\ref{lem1}. A specific choice of this testfunction is then made in order that
the inequality takes a particularly simple form (see lemma~\ref{lem2}) in terms of the harmonic energy of a
map into the coset $SL(n-2,\mr)/O(n-2,\mr)$. The inequality can be stated as saying that
the area is greater than or equal to a certain functional. By minimizing this functional over all possible
maps (see lemma~\ref{lem3}), we then obtain the desired inequality stated in 1) of Thm.~\ref{thm1}. The
rigidity statement 2) follows from the same reasoning.

\section{Special cases}

Before we give the proof, we consider
various special cases of the theorem for the sake of illustration.

\begin{enumerate}
\item
In $n=4$ dimensions, we evidently only have one angular momentum component $J=J_\pm$, so
the inequality reduces to that of~\cite{dain,dain1,dain2}. In this sense, our results generalize those
in four dimensions.

\item
In $n=5$ dimensions and topology type $\B \cong S^3$, the quantities $J_+,J_-$
correspond to different angular momentum components. By a making a suitable redefinition of
$\xi_1, \xi_2$ as in~\eqref{redef}, one can achieve that $J_+ = J_1$,
and $J_- = J_2$, so the inequality becomes
\ben\label{mpi}
A \ge 8\pi |J_1 J_2 |^\half \ .
\een
An illustrative example is provided by the Myers-Perry black hole~\cite{myers}. The area and angular momenta
of this solution can be written in parametric form as
\ben
A = 2\pi^2 r_0 \frac{(r_0^2+a_1^2)(r_0^2+a_2^2)}{r_0^2} \ , \qquad
J_i = \frac{(r_0^2+a_1^2)(r_0^2+a_2^2)\pi}{4r_0^2} \ a_i \ ,
\een
where $r_0>0$ is the location of the horizon in standard coordinates,
related to the surface gravity by $0\le \kappa = (r_0^4-a_1^2a_2^2)[r_0(r_0^2+a_1^2)(r_0^2+a_2^2)]^{-1}$.
One verifies explicitly that these expressions are compatible with the bound~\eqref{mpi},
with equality for the extremal black hole ($r_0^2 =|a_1 a_2|$).

\item
In $n=5$ dimensions and for the black ring topology type $\B \cong S^2 \times S^1$,
$J_+ = J_- = J_1$ is the angular momentum component in the $S^2$-direction, and the inequality reduces
to
\ben\label{bri}
A \ge 8\pi |J_1| \ .
\een
In particular, the lower bound is independent of the angular momentum $J_2$ in the $S^1$-direction, so
that the area remains bounded below no matter how large $J_2$ becomes. An illustrative example is provided
by the black ring solution~\cite{emparan} with two angular momenta~\cite{senkov}. In this case, the
area and $S^2$-angular momentum are given in parametric form by
\ben
A = \frac{32\pi^2k^3(1+\lambda+\nu)\lambda}{(y_0-1/y_0)(1-\nu)^2} \ , \quad
J_1 = \frac{4\pi k^3 \lambda \sqrt{\nu} \sqrt{(1+\nu)^2-\lambda^2}}{(1-\nu)^2(1-\lambda+\nu)}
\een
where $0\le \nu<1, 2\sqrt{\nu} \le \lambda<1+\nu$, and where $y_0=(2\nu)^{-1}(-\lambda+\sqrt{\lambda^2-4\nu})$
is the coordinate location of the horizon.
(Here we refer to the coordinates and notations of~\cite{senkov}). Some algebra confirms that these expressions are indeed compatible with the inequality~\eqref{bri}, with equality for the extremal black ring ($\lambda=2\sqrt{\nu}$).

\item
For $n=5$ dimensions and the lens space topology $\B \cong L(p,q)$,
we may achieve that $J_+=J_1$ and that $J_- = qJ_1 + pJ_2$
by a redefinition of the Killing fields as in eq.~\eqref{redef}. The inequality then states that
\ben
A \ge 8\pi |J_1(qJ_1+pJ_2)|^\half \ .
\een
We are not aware of the existence of regular stationary
black lenses, but it is possible to construct initial data for a dynamical asymptotically Kaluza
Klein (or asymptotically flat)
axisymmetric spacetime having a stable outer trapped surface $\B$ of
lens space topology. Our inequality would hence apply to such a spacetime.
\end{enumerate}

\section{Proof of thm.~\ref{thm1}}\label{proof}

The fist step of the proof is to take advantage of the stably outer marginally trapped condition
on $\B$, in combination with the field equations, $R_{\mu\nu}=\frac{2}{n-2} \Lambda g_{\mu\nu}$. There are several well-known and
essentially equivalent ways to do this, see e.g.~\cite{hayward,hayward1,hawkingellis} and many other references. Here we use
a method~\cite{racz} based on a special set of coordinates near $\H$. These ``Gaussian null coordinates'' $u,r,y^a$ are
defined as follows~\cite{moncrief,hiw}. First, we choose arbitrarily
local\footnote{Of course, it will take more than one patch to cover $\B$, but
the fields $\gamma, \beta, \alpha$ on $\B$ below in eq.~\eqref{gnc} are globally defined and
independent of the choice of coordinate systems.} coordinates $y^a$ on $\B$, and we Lie-transport them along the flow of $n$ to other places on $\H$, denoting by $u$ the affine parameter, and by $\B(u)$ the transported
 cross sections. Then, at each point of each $\B(u)$ we shoot off transversally null geodesics tangent to $l$
 with affine parameter $r$, where $g(n,l)=-1$. Then by definition $n = \partial/\partial u$, $l = \partial/\partial r$,
 and it can be shown that the metric then takes the Gaussian null form\footnote{Note that, if $\H$
 is an extremal stationary Killing horizon, then $n$ coincides with the stationary Killing field.
 For non-extremal stationary Killing horizons, $n$ differs from the Killing field.}
\ben\label{gnc}
g = -2 \d u (\d r  - r^2 \alpha \, \d u - r \beta_a \, \d y^a) + \gamma_{ab} \, \d y^a \d y^b \, ,
\een
where the function $\alpha$, the one-form $\beta = \beta_a \, \d y^a$, and the tensor field
$\gamma = \gamma_{ab} \, \d y^a \d y^b$ are invariantly defined tensor fields on each of the
copies $\B(u,r)$ of $\B=\B(0,0)$ of constant $r,u$. The horizon $\H$ is located by definition at $r=0$.
The Ricci tensor in these coordinates is given e.g. in~\cite{hiw}. The $ab$-components of the field equations
(see eq.~(82) in~\cite{hiw}) then give
\ben\label{ineq}
R(\gamma) - \half \beta_a \beta^a - D_a \beta^a = -2 \ \theta_n \theta_l- 2 \ {\mathscr L}_l \theta_n + 2 \ \Lambda \ge 0 \ ,
\een
where $R(\gamma)$ and $D$ are the scalar curvature and connection of $\gamma$ on $\B$, and where in the last step
we used the  conditions $\theta_n =0, {\mathscr L}_l \theta_n \le 0, \Lambda \ge 0$. (The conditions
$\theta_n \ge 0, \theta_l \le 0, {\mathscr L}_l \theta_n \le 0, \Lambda \ge 0$ would clearly also be sufficient.)

The next step is to use the fact that $g$, and hence $u,r,\alpha, \beta, \gamma$ are invariant under the
isometry group $G$. This is exploited in the following way. We first choose an axisymmetric testfunction
$\psi \in C^\infty(\B)$, multiply~\eqref{ineq} by $\psi^2$, and then integrate over $\B$ with
the integration element $\D S$ coming from $\gamma$. Then we get with $R \equiv R(\gamma)$:
\bena
0 &\le& \int_\B (R - \half \beta_a \beta^a - D_a \beta^a)\psi^2 \  \D S \non\\
&=& \int_\B (2\psi \beta^a D_a \psi - \half \beta_a \beta^a \psi^2 + R\psi^2) \ \D S \non\\
&=& \int_\B (2(\psi \beta_a N^a) N^b D_b \psi - \half (\psi N^a \beta_a)^2  - \half
(\gamma^{ab}-N^a N^b)\beta_a \beta_b \psi^2 + R\psi^2) \ \D S \non\\
&\le& \int_\B (2(N^b D_b \psi)^2 - \half
(\gamma^{ab}-N^a N^b)\beta_a \beta_b \psi^2 + R\psi^2) \ \D S \non\\
&=& \int_\B (2(D_b \psi) D^b \psi - \half
(\gamma^{ab}-N^a N^b)\beta_a \beta_b \psi^2 + R\psi^2) \ \D S \ .
\eena
Here, we denote by $N$ a unit normal\footnote{$N^a = D^a x/(D^b x D_b x)^\half$ is not defined at $x=\pm 1$
as $D_a x=0$ there,
but the expressions in the above integral are nevertheless well-defined.} tangent vector on $\B$ perpendicular to the $\xi_i$,
and in the third line we have used the basic inequality $2ab \le 2a^2 + \half b^2$. Thus,
we have shown the following lemma (compare lemma~1 of~\cite{dain}):

\begin{lemma}\label{lem1}
For any axisymmetric testfunction $\psi$ on $\B$ there holds
\ben
\int_\B (2(D_b \psi) D^b \psi - \half
(\gamma^{ab}-N^a N^b)\beta_a \beta_b \psi^2 + R\psi^2) \ \D S \ \ge 0 \ ,
\een
where $D,R,\D S$ are the intrinsic derivative operator, scalar curvature, and measure on $\B$.
\end{lemma}

To continue our discussion, it is useful to express the above inequality in terms of
the functions $f_{ij} = g(\xi_i, \xi_j)$ and certain potentials. These potentials are
defined as follows. First, we pass to the quotient $\hat \M = \M / G$. The global structure
of this quotient was described in detail in~\cite{hy}; here we only need to consider a
small open neighborhood of $\B$. In such a neighborhood, $\hat \M$ is simply parameterized
by the coordinates $u,r,x$ defined above. We complement these coordinates by $2\pi$-periodic
coordinates $\varphi^1, ..., \varphi^{n-3}$ in such a way that $\xi_i = \partial/\partial \varphi^i$.
Because $f_{ij}$ is positive definite (except at
the points $x=\pm 1$), $\hat \M$ inherits a Riemannian metric from the spacetime metric $g$.
We write this metric as $(\det f)^{-1} \ \D \hat s^2_3$, and the spacetime metric can then be written
in the usual Kaluza-Klein form as
\ben
g = f_{ij}(\D \varphi^i + \hat A^i)(\D \varphi^j + \hat A^j) + (\det f)^{-1} \ \D \hat s^2_3 \
\een
for 1-forms $\hat A^i$ on $\hat \M$. The condition $R_{\mu\nu} = \frac{2}{n-2} \Lambda g_{\mu\nu}$ implies the
``Maxwell equation''
\ben
\D \bigg( \det f \cdot f_{ij} \ \hat\star \ \D \hat A^j \bigg) = 0 \ ,
\een
implying locally the existence of potentials $\chi_i$ on $\hat \M$ satisfying
\ben
\D \chi_i = 2 \ \det f \cdot f_{ij} \ \hat\star \ \D \hat A^j \ \ .
\een
These so-called ``twist potentials'' are in fact defined globally~\cite{hy}, and can be viewed
as axisymmetric functions on $\M$, or alternatively on $\B$. In the latter case, they
are functions of $x$ only, and from~\eqref{komar} they satisfy
\ben\label{jump}
\frac{1}{8}(2\pi)^{n-4}\ \chi_i(x) \ \bigg|_{x=-1}^{x=+1} = J_i \ .
\een
So far, the coordinate $x \in [-1,+1]$ has only been an arbitrarily chosen parameter of the orbit space
$\hat \B = [-1,+1]$, but now we make a more specific choice. Our choice is fixed by
requiring the metric $\gamma$ on $\B$ to take the form
\ben\label{gammadef}
\gamma = \frac{\D x^2}{C^2 \det f} + f_{ij} \D \varphi^i \D \varphi^j \ ,
\een
where $C>0$ is some constant. We also set $\beta_i := i_{\xi_i} \beta$.
Then a straightforward calculation delivers the following expressions:
\bena
R &=& -\half C^2 f^{ij} \partial_x (\det f \ \partial_x f_{ij}) + \frac{1}{4} C^2 \det f \ f^{ij} f^{kl}
\partial_x f_{ik} \partial_x f_{jl} \non\\
&&- \frac{1}{4} C^2 \det f \  f^{ij} f^{kl}
\partial_x f_{ij} \partial_x f_{kl}
 - \half C^2 \det f \ \partial_x (f^{ij} \partial_x f_{ij}) \\
\beta_i &=&  C \partial_x \chi_i \\
\D S &=& C^{-1} \D x \prod_i \D \varphi^i \ ,
\eena
We insert this into lemma~\ref{lem1} with the choice $\psi = [(1-x^2)(\det f)^{-1}]^\half \in C^\infty(\B)$.
A longer but entirely straightforward calculation then gives the following beautifully simple
relation:
\begin{lemma}\label{lem2}
The functions $f_{ij}, \chi_j$ on $\B$ satisfy
\ben\label{ineq2}
0 \ge \int_{-1}^{+1} \left( \frac{1}{8} (1-x^2)\  {\rm Tr} (\Phi^{-1} \partial_x \Phi)^2  - \frac{1}{1-x^2} \right) \D x \ ,
\een
where the matrix $\Phi$ is defined by
\ben\label{phidef}
\Phi = \left(
\begin{matrix}
(\det f)^{-1} & -(\det f)^{-1} \chi_i \\
-(\det f)^{-1} \chi_i & f_{ij} + (\det f)^{-1} \chi_i \chi_j
\end{matrix}
\right) \quad .
\een
The matrix $\Phi$ is symmetric, positive definite, and $\det \Phi = 1$.
\end{lemma}
\noindent
{\bf Remark:}
As is well-known, the matrix $\Phi$ can be thought of as parameterizing the coset space
$SL(n-2,\mr)/O(n-2,\mr)$, and the quantity $\D s^2={\rm Tr}(\Phi^{-1} \D \Phi)^2$ gives
the natural Riemannian metric on that space, which can be shown to have negative curvature.
It is also well-known that Einstein's equations, with the symmetry group $G$ imposed,
reduce to that of a gravitating sigma-model on this space~\cite{maison} (for a concise
review of sigma-model approaches to higher dimensional gravity, see e.g.~\cite{Clement}). But
it is not clear to us why this implies that precisely the combination~\eqref{ineq2}
satisfies the beautifully simple inequality of lemma~\ref{lem2}!

\medskip

The lemma is our main technical tool. The final step in the proof consists of a variational
argument similar to that given in~\cite{dain1,dain2}. For this, we first note that, since
the metric $\gamma$ on $\B$ (see eq.~\eqref{gammadef}) is free of any type of conical singularity,
we must have the relation:
\ben\label{cdet01}
\frac{(1-x^2)^2}{\det f \cdot f_{ij} a^i_\pm a^j_\pm} \to  C^2 \quad \text{as $x \to \pm 1$.}
\een
We also note that, again from eq.~\eqref{gammadef}, the area $A$ of $\B$ is given by
\ben
A=2(2\pi)^{n-3} C^{-1} \ .
\een
We now simply add $2\log [A/2(2\pi)^{n-3}]$ to both sides of inequality~\eqref{ineq2}, taking into account~\eqref{cdet01}.
This gives, with $a^i(x):= \frac{1}{4} (1+x)^2 a_+^i + \frac{1}{4} (1-x)^2a_-^i$:
\bena\label{masterinq}
2 \log \frac{A}{2(2\pi)^{n-3}} &\ge& - \half x \log \frac{(1-x^2)^2}{\det f \cdot f_{ij} a^i a^j} \bigg|_{x=-1}^{x=+1} \non\\
&&
+ \int_{-1}^{+1} \left( \frac{1}{8} (1-x^2)\  {\rm Tr} (\Phi^{-1} \partial_x \Phi)^2  - \frac{1}{1-x^2} \right) \D x =: I[\Phi] \ .
\eena
The last equation can be viewed as the definition of a functional $I[\Phi]$ of $\Phi$.
(An alternative form of this functional, relating it to a functional studied in~\cite{dain,dain1}
in 4 dimensions, is given below in~\eqref{ialt}.)
We will obtain the inequality stated in thm.~\ref{thm1} by minimizing this functional.
However, in order to do this, we must say exactly in which class of matrix functions $\Phi$,
i.e. of the smooth, axisymmetric functions $f_{ij}, \chi_j$ on $\B$, we wish to vary $I[\Phi]$. For the
matrix $\Phi$ that arises from our metric $g$, we have the boundary conditions~\eqref{jump}
and~\eqref{bndycond}, and $f_{ij}$ must be non-singular and positive definite for $x \neq \pm 1$
with a one-dimensional kernel at $x = \pm 1$.
In fact, since $\gamma$ (see eq.~\eqref{gammadef}) is smooth, these conditions can be strengthened to the
requirements that
\ben\label{jump1}
\sigma := \log \frac{1-x^2}{\det f} \in C^\infty \ , \qquad \nu := \log \frac{1-x^2}{f_{ij} a^i a^j} \in C^\infty
\een
are smooth functions on $\B$, including $x = \pm 1$. Such axisymmetric $f_{ij}$, together with axisymmetric $\chi_i$ subject to the requirement~\eqref{jump}, constitute the class of $\Phi$ within which we seek a minimizer
of $I[\Phi]$. Our last lemma characterizes these minimizers:

\begin{lemma}\label{lem3}
Let $\Phi$ be smooth and axisymmetric such that conditions~\eqref{jump} and~\eqref{jump1} are satisfied,
and such that $I[\Phi]$ [cf.~\eqref{masterinq}] is finite. Then $I[\Phi] \ge I[\Phi_0]$, where $\Phi_0$
corresponds to the $f_{0ij}, \chi_{0j}$ of a near horizon geometry metric $g_0$ with
angular momenta $J_i$, see the appendix. Moreover, if $\Phi$ is a minimizer of
$I[\Phi]$, then $\Phi=\Phi_0$ where $\Phi_0$ corresponds to a near horizon metric.
\end{lemma}

\noindent
Before we give the proof of the last lemma, let us finish the proof of Thm.~1. The lemma
and eq.~\eqref{masterinq} immediately imply that
\ben
A \ge 2(2\pi)^{n-3} \e^{\half I[\Phi_0]} \ ,
\een
so we only need to calculate $I[\Phi_0]$.
As shown in~\cite{hi}, the near horizon geometries are characterized by a $\Phi_0$ satisfying the
equations
\ben\label{eom}
\partial_x[(1-x^2) \Phi^{-1}_0 \partial_x \Phi_0^{}] = 0 \ ,
\een
and the boundary conditions
 \ben\label{cdet1}
\frac{(1-x^2)^2}{\det f_0 \cdot f_{0 \ ij} a^i_\pm a^j_\pm} \to  C^2_0 \quad \text{as $x \to \pm 1$.}
\een
for some $C_0$ (not necessarily equal to $C$), as well as $f_{0 ij} a_\pm^i \to 0$ for $x \to \pm 1$ ,
and satisfying also $\chi_{0 i} \to \pm 4(2\pi)^{4-n} J_i$ for $x \to \pm 1$. The corresponding metrics $g_0$ that
follow from these conditions were determined in~\cite{hi}; they are reviewed for completeness in the appendix.
For us, the relevant consequence of the relations between the various parameters of the metric
$g_0$ stated in thm.~\ref{thm2} and~\eqref{relation1},~\eqref{constr}, is that:
\ben\label{ccond}
C_0 =
\frac{2(2\pi)^{(n-3)}}{8\pi |J_+ J_-|^\half} \ .
\een
(Note that this condition is true only for the near horizon metric $g_0$, but not necessarily the
metric $g$ that we started with.)
Furthermore, the equations of motion~\eqref{eom} for $\Phi_0$ and the boundary conditions
are seen to imply~\cite{hi} that
\ben
(1-x^2) \Phi_0^{-1} \partial_x \Phi_0^{} =2 \ S \left(
\begin{matrix}
1 & 0 & 0 & \dots & 0 \\
0 & 1 & 0 & \dots & 0 \\
0 & 0 & 0 & \dots & 0 \\
\vdots & & & & \vdots \\
0 & 0 & 0 & \dots & 0
\end{matrix}
\right)
S^{-1}
\een
for some constant unimodular matrix $S$. Hence, one immediately sees that the integral term
in $I[\Phi_0]$ [see eq.~\eqref{masterinq}] is precisely $=0$, which in view of~\eqref{ccond},~\eqref{cdet1} means
\ben
A \ge 2(2\pi)^{n-3} \ \e^{\half I[\Phi_0]} = 8\pi |J_+ J_-|^\half \ .
\een
This is the inequality claimed in part 1) of Thm.~1. To prove part 2), suppose first that
$g=g_0$ is a near horizon geometry. Then, again from \eqref{ccond}
and $A_0 = 2(2\pi)^{n-3} C_0^{-1}$, equality holds. Conversely, suppose that
the inequality in the theorem is saturated. Then the corresponding
matrix $\Phi$ must be a minimizer of $I[\Phi]$, and by lemma~\ref{lem3}, must
satisfy the equation~\eqref{eom} and must coincide with the $\Phi_0$ of a near horizon geometry. This implies
that the metric functions $\alpha, \beta, \gamma$ in~\eqref{gnc} must coincide with
those of a near horizon geometry, since these are determined by $\Phi$ as shown in~\cite{hi}.
This finishes the proof of Thm.~1. \qed

\medskip
\noindent
{\em Proof of lemma~\ref{lem3}:} The proof of this lemma has the same structure as that of lemma~4.1 of~\cite{dain1},
so we only outline the analogous steps and emphasize the new ideas that are needed.
The main point is that, if $\Phi$ is a minimum of $I[\Phi]$ [cf. eq.~\eqref{masterinq}],
then it is evident from the definition of $I[\Phi]$ that the matrix function $\Phi$ must satisfy the same Euler-Lagrange equations~\eqref{eom} that are satisfied by the $\Phi_0$ of a near horizon geometry. Since $\Phi$ must
satisfy boundary conditions~\eqref{jump} and~\eqref{bndycond}, it follows that a minimizer also satisfies the
same boundary conditions as a near horizon geometry. Hence it must be equal to a near horizon geometry.
The only gap in the argument is to show that $I[\Phi]$ actually {\em has} minimizers. Note that
the integral contribution to $I[\Phi]$ can be integrated further against $\D \varphi^1 \dots \D \varphi^{n-2}$,
and thereby essentially becomes the ``harmonic energy'' of  maps $\Phi$ from $\B$ into the negatively curved target space $SL(n-2,\mr)/O(n-2,\mr)$. The strategy is therefore to appeal to existence and uniqueness properties of harmonic maps into negatively curved target spaces~\cite{hildebrand}.
However, as stated, the result of~\cite{hildebrand} only applies to harmonic maps $\Phi: \Omega \to SL(n-2,\mr)/O(n-2,\mr)$
where $\Omega$ is a domain with non-empty boundary $\partial \Omega$, on which Dirichlet-type conditions are imposed.
On the other hand, our maps $\Phi$ are defined on $\B$, and as $x \to \pm 1$, behave in a singular way.
So, in order to apply the result of~\cite{hildebrand}, one has to get around these problems. This was
explained in \cite{dain1} for the case $n=4$, in which the target space is isometric to the 2-dimensional hyperbolic space
$SL(2,\mr)/O(2,\mr) \cong {\mathbb H}$. The same type of argument also applies in the present context.

We first pick an arbitrary $\Phi_0$ corresponding to a near horizon geometry in the class of functions
satisfying eq.~\eqref{jump1}, with the same values of the $J_i$.
Then we consider a domain $\Omega_I \subset \B$ which is a small neighborhood
$$\Omega_I=\{p \in \B \mid |x(p) - 1|<\e^{-(\log \epsilon)^2}  \ \ {\rm or} \ \ |x(p)+1|<\e^{-(\log \epsilon)^2}\}$$
 of the points characterized by
$x = \pm 1$ where the map $\Phi$ behaves in a singular way. We take a suitable cutoff function $\psi_\epsilon$
which is equal to 1 on $\Omega_I$, and which is equal to 0 on a region
$$\Omega_{III}=\{p \in \B \mid |x(p)-1|>\epsilon \ \ {\rm and} \ \ |x(p)+1|> \epsilon\}$$
and which interpolates
between $0$ and $1$ in the remaining intermediate annular region $\Omega_{II}$. One then
considers an interpolation (constructed using $\psi_\epsilon$)
$\Phi_\epsilon$ so that $\Phi_\epsilon = \Phi_0$ on $\Omega_I$, and so that $\Phi_\epsilon = \Phi$ on
$\Omega_{III}$. The explicit form of this interpolation will be described below.
Having defined the interpolation, one defines $I_\epsilon[\Phi]:=I[\Phi_\epsilon]$. Now on $\Omega_{IV} = \Omega_{III} \cup \Omega_{II}$,
the function $\Phi_\epsilon: \Omega_{IV} \to SL(n-2,\mr)/O(n-2,\mr)$ by construction has the same boundary
values as $\Phi_0$, and so on $\Omega_{IV}$ cannot have smaller harmonic energy by the result of~\cite{hildebrand},
i.e. $I_\epsilon |_{\Omega_{IV}}$ (we mean restriction to maps defined on $\Omega_{IV}$ with
prescribed boundary conditions), is not smaller than $I_\epsilon |_{\Omega_{IV}}$ evaluated on $\Phi_0$. On the other hand, on $\Omega_I$, we have by definition $I_\epsilon |_{\Omega_I} [\Phi] = I_\epsilon |_{\Omega_I}[\Phi_0]$, so one
obtains that $I_\epsilon[\Phi] \ge I_\epsilon[\Phi_0] = I[\Phi_0]$. So, one is done if one can show that
$\lim_{\epsilon \to 0} I_\epsilon = I$. This part requires a special choice of the cutoff function--we may
chose the same one as in eq.~(71) of~\cite{dain1}--and a special interpolation.

To choose the interpolation, we parameterize the coset
space (i.e. matrices $\Phi$) by $f_{ij}, \chi_i$ as above. But we further parameterize $f_{ij}$ by
$\nu, \sigma$ (defined above in~\eqref{jump1}), and by new quantities $c^{ij}, b^i$:
\ben
f^{ij} = \left( \frac{\e^{-\nu}}{1-x^2} + c_{kl} b^k b^l \right) a^i a^j + 2a^{(i} b^{j)} + c^{ij} \ .
\een
The functions $a^i$ were defined above before eq.~\eqref{masterinq}.
The quantities $b^i$ and $c^{ij}$ can be viewed as linear maps on covectors $v_i$. On covectors which satisfy $a^i v_i = 0$, the bilinear form $c^{ij}$ is positive-definite and non-degenerate, everywhere on $\B$,
including $x=\pm 1$. To make $c^{ij}, b^i$ unique, we choose an arbitrary but fixed covector
function $a_i$ such that $a_i a^i = 1$, and we demand that $c^{ij}a_j = 0 = b^j a_j$.
It follows from these definitions that there is a $c_{ij}$ such
that $c_{ij} a^j = 0$ and such that $c_{ij} c^{jk} v_k = v_i$ for all
$v_i$ satisfying $v_i a^i = 0$. We denote by $c$ the determinant of $c^{ij}$ when restricted to $v_i$'s
of this form, and we also denote $b_i = c_{ij} b^j$. Combining these formulae with~\eqref{jump1},
we see that $c$ is constrained to be of the form $c = \e^{\nu-\sigma}$.
Thus, we parameterize $\Phi$ by the quantities $\sigma, \nu, b^i, c^{ij}, \chi_i$, all of which are {\em smooth} on $\B$, including at $x = \pm 1$, and axi-symmetric. We can then simply linearly interpolate
\bena
&&\sigma_\epsilon = \psi_\epsilon \sigma_0 + (1-\psi_\epsilon) \sigma \ ,
\quad
b_\epsilon^i = \psi_\epsilon b_0^i + (1-\psi_\epsilon) b^i_{} \ , \\
&&\nu_\epsilon = \psi_\epsilon \nu_0 + (1-\psi_\epsilon) \nu \ ,
\quad
\chi_{\epsilon i} = \psi_\epsilon \chi_{0i} + (1-\psi_\epsilon) \chi_i
\eena
and we interpolate
\ben
c^{ij}_\epsilon =
[\exp(\psi_\epsilon A_0) \exp((1-\psi_\epsilon)A)]^{ij}
\een
where $A_0^{ij}, A_{}^{ij}$ are the matrix logarithms of $c^{ij}_0, c^{ij}_{}$.
The last definition ensures that $c_\epsilon = \e^{\nu_\epsilon-\sigma_\epsilon}$.
We denote the interpolated matrix parameterized by $\sigma_\epsilon, \nu_\epsilon, b^i_\epsilon, c^{ij}_\epsilon,
\chi_{\epsilon i}$
by $\Phi_\epsilon$. Our parameters have been chosen in such a way that the functional $I[\Phi]$ takes a form
in which we can analyze it in the same way as done in~\cite{dain1}. Namely, a longer, but entirely straightforward,
calculation shows that $I[\Phi]$ is given by:
\bena\label{ialt}
&&I[\Phi] = \\
&& \frac{1}{8} \int_\B \left( -8 + 4 \ \nu + 4 \ \sigma + (\partial_\theta \nu)^2 + (\partial_\theta \sigma)^2 + 2 \ \frac{a^i a^j
(\partial_\theta \chi_i)\partial_\theta \chi_j}{\sin^4 \theta \ \e^{-\nu-\sigma}} \right) \D S_0 + \non\\
&& \frac{1}{8} \int_\B \bigg(
-b^k b_k \bigg[ 2\frac{\e^{-\nu}}{\sin^2 \theta}  +    b^i b_i\bigg] (\partial_\theta(\e^\nu \sin^2 \theta))^2 \non\\
&&\hspace{1cm}
 -4\ \bigg[ 1 + b^j b_j \e^\nu \sin^2 \theta \bigg]^2 \bigg[ \frac{\cos \theta}{\sin \theta} + \partial_\theta
 \nu -  a^l \partial_\theta a_l -  a^l \partial_\theta b_l \bigg]
 (a_i \partial_\theta a^i + b_i \partial_\theta a^i) \non\\
&&\hspace{1cm}
- \bigg[ \frac{\e^{-\nu}}{\sin^2 \theta}  +    b^n b_n\bigg]^2 \partial_\theta(a^i a^k) \partial_\theta (a^j a^l)
(\e^\nu \sin^2 \theta \ a_i a_k - 2a_{(i} b_{k)})(\e^\nu \sin^2 \theta \ a_j a_l - 2a_{(j} b_{l)}) \non\\
&&\hspace{1cm} +8 \  (1 +    b^n b_n \e^\nu \sin^2 \theta ) \bigg[
\frac{\cos \theta}{\sin \theta} + \half \partial_\theta \nu + a^k \partial_\theta a_k - a^k \partial_\theta b_k
\bigg] \non\\
&&\hspace{1cm}
\cdot \ \ \bigg[
\e^\nu \sin^2 \theta \ b^l \partial_\theta a_l - 2 \partial_\theta (\e^\nu \sin^2 \theta) b^l b_l - 2 \e^\nu
\sin^2 \theta \ a^j b^l \partial_\theta(a_j b_l) - b_l \partial_\theta a^l
\bigg] \non\\
&&\hspace{1cm}
+2 \ \bigg[ \frac{\e^{-\nu}}{\sin^2 \theta}  +    b^n b_n\bigg] c^{kl} \bigg[
\e^\nu \sin^2 \theta \ \partial_\theta a_k - 2a^i \partial_\theta(\e^\nu \sin^2 \theta \ a_i b_k)
- c_{ik} \partial_\theta a^i
\bigg] \non\\
&&\hspace{1cm}
\cdot \ \
\bigg[
\e^\nu \sin^2 \theta \ \partial_\theta a_l - 2a^j \partial_\theta(\e^\nu \sin^2 \theta \ a_j b_l)
- c_{jl} \partial_\theta a^j
\bigg]\non\\
&&\hspace{1cm}
+(c^{ij} + 2b^{(i} a^{j)})(c^{kl} + 2b^{(k} a^{l)})
\partial_\theta\bigg[c_{ik} + \e^\nu \sin^2 \theta  \ b_i b_k + \e^\nu \sin^2 \theta \ a_i a_k
-2\e^\nu \sin^2 \theta \ a_{(i} b_{k)} \bigg]\non\\
&&\hspace{1cm}
\cdot \ \
\partial_\theta\bigg[c_{jl} + \e^\nu \sin^2 \theta  \ b_j b_l+ \e^\nu \sin^2 \theta \ a_j a_l
-2\e^\nu \sin^2 \theta \ a_{(j} b_{l)}\bigg] \non\\
&&\hspace{1cm}
+ 2 \ \frac{\e^{-\sigma}}{\sin^2 \theta} b_i b^i a^j a^k (\partial_\theta \chi_j) \partial_\theta \chi_k
+ 4 \ \frac{\e^{-\sigma}}{\sin^2 \theta} b^j a^k (\partial_\theta \chi_j) \partial_\theta \chi_k
+ 2 \ \frac{\e^{-\sigma}}{\sin^2 \theta} c^{jk} (\partial_\theta \chi_j) \partial_\theta \chi_k
\bigg) \D S_0 \ .\non
\eena
Here, we have defined $x = \cos \theta$, and also $\D S_0 = (2\pi)^{3-n} \ \sin \theta \D \theta \prod
\D \varphi^j$, which coincides up to a constant factor with the measure $\D S$ on $\B$ coming from the metric. The terms have been grouped in the following way. The first
integral, which we call $M[\Phi]$, is analogous to the ``mass functional'' of~\cite{dain1}: In $n=4$, we have that $\nu = \sigma$
is equal to the quantity $\zeta$ in that reference, $a^1 = a_1 = 1$,  $\chi_1$ is equal to the quantity $\omega$
in that reference, and $\B = S^2$. Then $M[\Phi]$ is actually exactly equal to the mass functional defined
in that reference. Also in higher dimensions, the integral has a completely analogous structure and can therefore be handled in exactly in the same way as in~\cite{dain1} to show that $M_\epsilon[\Phi] \equiv M[\Phi_\epsilon] \to M[\Phi]$ as $\epsilon \to 0$. The second integral, which we call $M'[\Phi]$, has only terms that are manifestly regular at $\theta = 0, \pi$ (here we also use that $\partial_\theta \chi_i = O(\sin \theta)$ due to~\eqref{jump},
and that $\partial_\theta a^i = O(\sin^2 \theta)$). It not present in $n=4$, but because the integrand is regular, it is straightforward to see that it converges,
$M'_\epsilon[\Phi] \to M'[\Phi]$. Because the arguments are very similar, and no new ideas are required,
we will not give the details here.
\qed

\vspace{2cm}
\noindent
{\bf Conventions:} Our conventions for the signature of the metric and the
curvature tensor are the same as in Wald's textbook~\cite{wald}. Our convention for the
Hodge dual $\star$ is such that $\star^2 = -1$.

\vspace{1cm}
\noindent
{\bf Acknowledgements:} It is a pleasure to thank S.~Dain, A.~Ishibashi, and J.~Lucietti
for their comments on the first draft of this paper.
\appendix

\section{Near horizon geometries in $n$ dimensions}

The near horizon geometry of an extremal $n$-dimensional vacuum stationary black hole with
isometry group\footnote{Note that the notion of near horizon geometry is defined for
any extremal black hole with Killing horizon, which does not have to possess the isometry
group $G$. The ones we discuss here thus special, ``codimension-1'', near horizon geometries.} $\mr \times G$ is defined by a certain scaling limit of the metric near the horizon.
This scaling limit is a new Ricci-flat metric and automatically possesses an enhanced symmetry
group of $O(2,1) \times G$, see~\cite{bardeen,lucietti}.
The classification of all such geometries was achieved first in $n=4,5$ in~\cite{lucietti1,lucietti2}. In $n=4$,
it had been known for a long time and is also sometimes referred to as the ``extremal Kerr-throat''.
The generalization to $n$-dimensions, was performed in~\cite{hi}. The method of proof in~\cite{hi} shows that
the matrix $\Phi_0$ formed from the quantities $f_{0 ij}, \chi_{0 i}$ satisfies the equations \eqref{eom}.
Combining this with the boundary conditions, one arrives at the following classification~\cite{hi}:

\begin{thm}\label{thm2}
All non-static near horizon metrics
are parameterized by the real parameters $c_\pm, \mu_i, s_{Ii}$,
and the integers $a_\pm^i$ where $I=0,\dots,n-5$ and $i=1, \dots, n-3$,
and ${\rm g.c.d.}(a_\pm^i) = 1$.
The explicit form of the near horizon metric in terms of these parameters is
\bena\label{NH}
g_0 &=& \e^{-\lambda} (2\d u \d R - C^2_0 R^2 \d u^2 + C^{-2}_0 \, \d \theta^2) +
\e^{+\lambda} \Bigg\{ (c_+-c_-)^2 (\sin^2 \theta) \, \Omega^2 \non\\
&& +(1+\cos \theta)^2 c_+^2 \sum_I \left( \omega_I - \frac{s_I \cdot a_+}{\mu \cdot a_+} \Omega \right)^2
+(1-\cos \theta)^2 c_-^2 \sum_I \left( \omega_I - \frac{s_I \cdot a_-}{\mu \cdot a_-} \Omega \right)^2\non\\
&& +  \frac{c_\pm^2\, \sin^2 \theta}{(\mu \cdot a_\pm)^2} \sum_{I < J} \Big(
(s_I \cdot a_\pm) \omega_J - (s_J \cdot a_\pm)\omega_I \Big)^2
\Bigg\} \, .
%
\eena
Here, the sums run over $I,J$ from $0, \dots, n-5$, the function
$\lambda(\theta)$ is given by
\ben
\exp[-\lambda(\theta)] = c_+^2(1+\cos \theta)^2 + c_-^2(1-\cos \theta)^2 + \frac{c_\pm^2 \sin^2 \theta}{(\mu\cdot a_\pm)^2} \sum_I (s_{I}\cdot a_{\pm})^2 \, ,
\een
$C_0$ is given by $C_0 = 4c^2_\pm[(c_+-c_-)(\mu \cdot a_\pm)]^{-1}$, and we have defined  the 1-forms
\bena
\Omega(R) &=& \mu \cdot \d \varphi + 4C_0R\frac{c_+c_-}{c_+ -c_-} \d u \\
\omega_I(R) &=& s_{I} \cdot \d \varphi +
\frac{r}{2} \, C^2_0 (s_{I} \cdot a_+ + s_I \cdot a_-)  \, \d u \, .
\eena
We are also using the shorthand notations such as $s_{Ii} a^i_+ = s_I \cdot a_+$, or
$\mu \cdot \d \varphi = \mu_i \d \varphi^i$, etc.
The parameters are subject to the constraints $\mu \cdot a_\pm \neq 0$ and
\ben\label{constr}
\frac{c_+^2}{\mu \cdot a_+} = \frac{c_-^2}{\mu \cdot a_-}
\, ,
\quad
\frac{c_+ (s_{I} \cdot a_+)}{\mu \cdot  a_+} = \frac{c_- (s_{I} \cdot a_-)}{ \mu \cdot  a_-}
\, ,
\quad
\pm 1 = (c_+ - c_-) \, \epsilon^{ijk \dots m} s_{0i} s_{1j} s_{2k} \cdots \mu_m
\een
but they are otherwise free.  When writing ``$\pm$'', we mean that the
formulae hold for both signs.
\end{thm}
\noindent
{\bf Remarks:}
(1) The part $2 \d u \d R - C^2_0 R^2 \d u^2$ of the metric is $AdS_2$-space with curvature $C^2_0$. This is the cause for the
enhanced symmetry group of $O(2,1) \times U(1)^{n-3}$, as explained in more detail in~\cite{lucietti1}.

(2) The coordinate $\theta \in [0,\pi]$ is related to the coordinate $x$ in the previous section by
$x = \cos \theta$. The relation between $r$ and $R$ is more complicated and is explained in~\cite{hi}.

\medskip
\noindent
The constants $\mu_i, c_\pm, a^i_\pm$ are directly related to the
horizon area by
\ben
A_0 = 2(2\pi)^{n-3} C_0^{-1}
\een
and we also have
\ben\label{relation1}
8\pi J_i  = (2\pi)^{n-3} \, \frac{c_+ - c_-}{2c_-c_+} \mu_i \, .
\een

\end{document}